\newcommand{\beq}{\begin{equation}}
\newcommand{\eeq}{\end{equation}}
\newcommand{\beqns}{\begin{equation}}
\newcommand{\eeqns}{\end{equation}}
\newcommand{\beqar}{\begin{eqnarray}}
\newcommand{\bs}{\begin{eqnarray*}}
\newcommand{\eeqar}{\end{eqnarray}}
\newcommand{\es}{\end{eqnarray*}}
\newcommand{\beqml}{\begin{mathletters}}
\newcommand{\eeqml}{\end{mathletters}}
\newcommand{\Ptilde}{\tilde{P}}

\documentstyle[aps,preprint]{revtex}
\begin{document}
\draft
\title{Exact closed form of the return probability on the Bethe lattice}
\author{Achille Giacometti}
\address{Institut f\"{u}r Festk\"{o}rperforschung
der Kernforschungsanlange, \\
Postfach 1913, D-52425,  J\"{u}lich, Germany}
\date{\today}
\maketitle
%\addtolength{\baselineskip}{\baselineskip}

%%%%%%%%%%%%%%%%%%%%% Begin Abstract %%%%%%%%%%%%%%%%%%%%%%%%%%%%%%
\begin{abstract}
An exact closed form solution for the return probability of a random walk
on the Bethe lattice is given. The long-time asymptotic form
confirms a previously known expression. It is however shown that
this exact result reduces to the proper expression when
the Bethe lattice degenerates on a line, unlike the asymptotic
result which is singular. This is shown to be an artefact of the
asymptotic expansion. The density of states is also calculated.
\end{abstract}
%
%%%%%%%%%%%%%%%%%%%%%% End Abstract %%%%%%%%%%%%%%%%%%%%%%%%%%%%%%
\newpage
\narrowtext
Beside being an interesting type of graph {\it per se}, the Bethe lattice
(BL) is also reckoned as paradigm of a lattice in the
limit of high dimensionality.
A BL (see Fig.\ref{fig1}) is usually defined as a set of sites
connected by bonds,
such that each site has the same coordination number and
there are no closed loops. It differs from the so called
Cayley tree on the fact that the complication
arising by the boundary conditions is neglected \cite{Baxter}.

The problem of the random walk on a BL it is not new \cite{HS,Cassi}.
However up to now only asymptotic expressions were given.
One of the surprising features of these asymptotic expressions
was the difficulty arising in the interpretation of
the result in the limit when the BL collapses into a line.

Main aim of the present note is to derive an exact closed form
solution which on one hand confirms previous asymptotic results,
but on the other hand has the proper limit form
when the BL reduces to a one-dimensional lattice, thus confirming
the exactness of the asymptotic procedure and solving
the aforementioned interpretation puzzle.
Moreover we will provide an alternative solution approach
with respect to the previous investigations.

Let us start from the general master equation on the lattice:
\beqar \label{ME}
P_{x,0}(t+1) &=& P_{x,0}(t) + \sum_{y(x)} [w_{x,y} P_{y,0}(t)
                -w_{y,x} P_{x,0}(t)]
\eeqar
where $P_{x,0}(t)$ is the probability density of being at site
$x$ at time $t$ having started from site $0$ at the initial time $t_0=0$.
The notation $y(x)$ means that the sum is restricted to the
nearest neighbours $y$ of $x$.
In the Bethe lattice case, $w_{x,y}=1/z$ where $z$ is the coordination
number of the lattice.

It is convenient to introduce the generating function of $P_{x,0}(t)$
(Green function):
\beqar \label{Generating}
\Ptilde_{x,0}(\lambda) &=& \sum_{t=0}^{+\infty} \lambda^t P_{x,0}(t)
\eeqar
The fact that all points belonging to the same shell are topologically
equivalent, allows to map the solution for the Bethe lattice onto
the solution of a one-dimensional lattice with a defect.
Therefore the Green equation takes on the Bethe lattice the form:
\beqml
\beqar
\Ptilde_{0,0}(\lambda) &=& \lambda \Ptilde_{1,0}(\lambda) +1
\label{GF:1} \\
\Ptilde_{n,0}(\lambda) &=& \frac{\lambda}{z} \Ptilde_{n-1,0}(\lambda)+
\frac{\lambda(z-1)}{z} \Ptilde_{n+1,0}(\lambda)
\label{GF:2}
\eeqar
\eeqml
for the $0-$th and $n-$th shell respectively. Here $P_{n,0}(t)$
refers then to the probability of being in the $n-$th shell at time
$t$ having started from the seed $0$.

The solution of eq. (\ref{GF:1},\ref{GF:2}) is considerably simplified by
noting that the ratio $\Ptilde_{n+1,0}(\lambda)/\Ptilde_{n,0}(\lambda)$
is {\it independent} on $n$ due to the homogeneity of the lattice
and the particular boundary conditions \cite{Thorpe}.
It is then a simple matter to solve the quadratic equation coming
from (\ref{GF:2}) and substitute the root,
which has finite value in the $\lambda \rightarrow 0$ limit,
into (\ref{GF:1}) with the result:
\beqar \label{GF_0}
\Ptilde_{0,0}(\lambda) &=& \frac{2(z-1)/z}{(z-2)/z+
\sqrt{1-\frac{4\lambda^2(z-1)}{z^2}}}
\eeqar
This result was previously obtained by Cassi \cite{Cassi} by a different
procedure.

It is important to notice that for $z=2$ this expression reduces to the
well known \cite{BN} result of the generating
function for the one dimensional lattice. It is also worth mentioning
that since the critical value for the fugacity
$\lambda_c=z/2\sqrt{z-1}>1$ for $z\ge3$ the generating function
$\Ptilde_{0,0}(\lambda)$ is always real and finite for $\lambda \le 1$
and therefore a random walk on the Bethe lattice {\it cannot } be critical.

Upon serie expansion of eq.(\ref{GF_0}) and
using the definition (\ref{Generating}), one gets after some algebra:
\beqar \label{Intermediate}
P_{0,0} (2t) &=& \frac{(z-1)}{z} (\frac{\sqrt{z-1}}{z})^{2t}
\sum_{p=0}^{+\infty} \frac{(2p+2t)!}{(p+t+1)!(p+t)!} (\frac{z-1}{z^2})^p
\eeqar
and $P_{0,0}(2t+1)=0$ for $t>0$ with $P_{0,0}(0)=1$.
After some manipulations this expression can be cast in the following
closed form:
\beqar \label{Exact}
P_{0,0}(2t) &=& \frac{(z-1)}{z} (\frac{\sqrt{z-1}}{z})^{2t} \frac{\Gamma(2t+1)}
{\Gamma(t+2) \Gamma(t+1)} \;_2F_1(t+1/2,1,t+2,4(z-1)/z^2)
\eeqar
where $\Gamma(t)$ is the gamma function and $_2F_1(\alpha,\beta,\gamma,z)$
is the Gauss hypergeometric function \cite{AS}.

For large $t$ one can use the property \cite{note}
\beqar \label{property1}
_2F_1(\alpha,\beta,\beta,z) &=& (1-z)^{-\alpha}
\eeqar
valid for arbitrary $\beta$ and the Stirling approximation
for the gamma function \cite{AS} to find, at the leading order in $t>>1$:
\beqar \label{Asymptitic}
P_{0,0}(t) &\stackrel{t>>1}{\sim}& \frac{2^{3/2}  z(z-1)}{\sqrt{\pi}
(z-2)^2} t^{-3/2} \exp(-t \ln (\frac{z}{2 \sqrt{z-1}}))
\eeqar
which confirms the asymptotic result
derived in ref.\cite{Cassi}.  Note that in the limit
$z \rightarrow 2$ (when the Bethe lattice degenerates on a line)
the expected asymptotic behaviour $P_{0,0}(t) \sim t^{-1/2}$
is {\it not} recovered both because the prefactor becomes singular
and because the (universal) power law is not the correct one.

We are now in the position to show that this is an artefact of
the asymptotic expansion stemming from the fact
that the limit $t \rightarrow \infty$ and $z \rightarrow 2$ do
not commute. Indeed if we set $z=2$ in the
exact result (\ref{Exact}) and use the fact that \cite{AS}:
\beqar \label{property2}
_2 F_1 (\alpha,\beta,\gamma,1) &=& \frac{\Gamma(\gamma)
\Gamma(\gamma-\alpha-\beta)}{\Gamma(\gamma-\alpha)\Gamma(\gamma-\beta)}
\eeqar
we get:
\beqar \label{one-dimensional}
P_{0,0} (t) &=& (1/2)^t \frac{\Gamma(t+1)}{\Gamma(t/2+1) \Gamma(t/2+1)}
\eeqar
which is the well-known result of the one-dimensional case \cite{BN}.
This latter result could in fact be derived by starting from
eq.(\ref{GF_0}) for $z=2$ and using a procedure similar
to the one employed here.

It is also possible to compute the density of states, associated
to the master equation (\ref{ME}), on the BL. This was
not previously calculated. Indeed using the transformation
procedure between discrete and continuum times described
in ref \cite{GMN}, it is not hard to see that the
the Laplace transform of the return probability is
related to the generating function $\Ptilde_{0,0}(\lambda)$ through:
\beqar \label{mapping}
\Ptilde_{0,0}(\omega) &=& \lambda
\Ptilde_{0,0}(\lambda)|_{\lambda=1/(1+\omega)}
\eeqar
where $\omega$ is the Laplace variable conjugate of time.
Then, using eq.(\ref{GF_0}) and (\ref{mapping}), one obtains:
\beqar \label{Laplace}
\Ptilde_{0,0}(\omega) &=& \frac{2(z-1)/z}{\frac{(z-2)}{z}(1+\omega)+
\sqrt{\omega^2+2\omega+(z-2)^2/z^2}}
\eeqar
The density of state $\rho(\epsilon)$ is then well known \cite{HK}
to be given by the analytical continuation:
\beqar \label{Analitical}
\rho(\epsilon) &=& - \frac{1}{\pi} \mbox{Im} \Ptilde_{0,0}(-\epsilon +i 0^{+})
\eeqar
Then, in the present case, the result of the analytical continuation is:
\beqar \label{DOS}
\rho(\epsilon) &=& \left\{ \begin{array}{ll}
             \frac{z}{2 \pi} \frac{\sqrt{2 \epsilon -\epsilon^2
	          -(z-2)^2/z^2}}{2 \epsilon-\epsilon^2} \;\;
              \mbox{if $2 \epsilon-\epsilon^2 -(z-2)^2/z^2 >0$} \\
              \;\;\;\;\; 0 \;\; \mbox{otherwise}
              \end{array}
               \right.
\eeqar
which for $z=2$ reduces to the well known formula of
the density of states of a one-dimensional lattice \cite{Bernasconi}.
Fig.\ref{fig2} shows the comparison between the case of the
one-dimensional lattice ($z=2$) and the BL ($z=3,4$).

Summarizing we presented an exact closed form solution
for the return probability and the density of states
on the BL. Previous asymptotic results were confirmed and
explained in terms of this solution.
Although other relevant quantities, beside the ones presented here,
could, in principle, be obtained, the required algebra becomes
rapidly very involved. This is beyond the purposes of the present
work whose main objective was the confirmation of the asymptotic results,
along with the removal of the inconsistency contained in them.
As a by-product of our investigation we also
presented an alternative simplified solution procedure
with respect to the previous two approaches.

This work was supported by the {\it Human Capital and Mobility} program
under the contract ERB4001GT932058. I wish to thank Amos Maritan,
Klaus Kehr and Flavio Seno for a critical reading of the manuscript.
I am especially grateful to Davide Cassi and Sofia Regina for
their many constructive criticisms and suggestions and
for having pointed out an error in the original draft of the manuscript.
%%%%%%%%%%%%%%%%%%%%%% Bibliography %%%%%%%%%%%%%%%%%%%%%%%%%%%%%%%%%%%%

%%%%%%%%%%%%%%%%%%%%%%%% Figures %%%%%%%%%%%%%%%%%%%%%%%%%%%%%%%%%%%%%%%
%Fig 1%
\begin{figure}
\caption{Example of a Bethe lattice with coordination number $z=3$.
Sites indicated with the same numbers $n=0,1,2,...$ belong to the same shell.}
\label{fig1}
\end{figure}
%Fig 2%
\begin{figure}
\caption{Density of states associated to the master equation in the case
of a one-dimensional lattice ($z=2$) and of a Bethe lattice ($z=3,4$).}
\label{fig2}
\end{figure}
\end{document}